\documentclass[12pt]{article}

\usepackage{amstext,amsmath,amssymb,amsfonts,bbm}
\usepackage[latin1]{inputenc}
\usepackage{epsfig}
\usepackage{hyperref}
\usepackage{amsthm}
\usepackage{subfigure}
\usepackage{color}
\usepackage{multirow}
\newcommand{\bea}{\begin{eqnarray}}	
\newcommand{\eea}{\end{eqnarray}}
\newcommand{\cG}{{\cal G}}
\newcommand{\cV}{{\cal V}}
\newcommand{\cC}{{\cal C}}

\newcommand{\cB}{{\cal B}}

\newtheorem{lemma}{Lemma}

\newtheorem{remark}{Remark}

\newtheorem{theorem}{Theorem}
\newtheorem{proposition}{Proposition}

\begin{document}

\title{Lost in Translation: Topological Singularities in Group Field Theory}

\author{Razvan Gurau\footnote{Perimeter Institute for Theoretical Physics Waterloo, ON, N2L 2Y5, Canada } }

\maketitle

\begin{abstract}
\noindent
Random matrix models generalize to Group Field Theories 
(GFT) whose Feynman graphs are dual to gluings of higher dimensional
simplices. It is generally assumed that GFT graphs are always dual to 
pseudo manifolds. In this paper we prove that already in dimension three (and 
in all higher dimensions), this is {\it not true} due to subtle differences 
between simplicial complexes and gluings dual to GFT graphs. 

We prove however that, fortunately, the recently introduced ``co-lored'' GFT models 
\cite{color} do not suffer from this problem and {\it only} ge-nerate 
graphs dual to pseudo manifolds in any dimension.
\end{abstract}

\section{Introduction: Group Field Theory}
\label{sec:intro}

Group Field Theories (GFT) \cite{GFT,laurentgft} are  
quantum field theories over group manifolds. They generalize random matrix models and 
random tensor models \cite{mmgravity,ambj3dqg} (see also \cite{sasa1,sasa2}). 
GFT's arise naturally in several discrete approaches to quantum gravity, 
like Regge calculus \cite{williams}, dynamical triangulations \cite{DT} or 
spin foam models \cite{SF} (see \cite{libro} for further details).

The Feynman graphs of GFT are built from vertices encoding the 
connectivity dual to a $n$ simplex, and propagators encoding 
the connectivity dual to the gluing of $n$ simplices along boundary $(n-1)$
simplices. A graph is dual to a ``gluing of simplices'' 
yielding some $n$ dimensional topological space, and GFT's
generalize the familiar matrix models \cite{mm} to a theory
of random higher dimensional topological spaces.

In discrete approaches to quantum gravity \cite{libro,danielegft}, a gluing 
of simplices is interpreted as a space-time background making GFT a combinatorial, 
background independent theory, whose perturbative development generates space-times.
This is further supported as, for the simplest choice of vertex and propagator, the Feynman 
amplitude of a graph reproduces the partition function of a BF theory discretized 
on the gluing of simplices \cite{GFT,newmo}\footnote{In algebraic combinatorics this lead to
new topologycal invariants \cite{TV} and advances on 
the volume conjecture \cite{malek1,malek2}.}. BF theory becomes 
Einstein gravity after implementing the Plebanski constraints and it is natural to suppose that a 
some more involved GFT model will reproduces the partition function 
of the latter. Working at the level of individual GFT graphs (spin foams), one has a good control 
over the constraints, and their implementation leads to several alternative propositions
(\cite{newmo1, newmo2} or \cite{newmo3, newmo4}) of vertex kernels. The semiclassical limit 
\cite{semicl1, semicl2} of these models has been analyzed with encouraging results.
Alternatively, one can try to implement the constraints directly at the level of the action
(\cite{gftquantgeom,gftnoncom} or \cite{quantugeom2}) or include 
matter fields \cite{matter1,matter2}. Recently GFT's and spin foams have been 
adapted to the study of loop quantum cosmology \cite{Ashtekar:2009dn,Ashtekar:2010ve}.

Irrespective of the particularities of the model, the fundamental question in all 
discrete approaches to quantum gravity is ``to sum or not to sum?''. According
to the answer to this question one distinguishes several possibilities.
Spin foam models sum over all metrics at fixed triangulation, dynamical triangulations sum over 
subclasses of triangulations\footnote{With metric fixed for a given triangulation.} at fixed topology,
while GFT's sum over everything. The weights of different topologies, 
triangulations and metrics are completely fixed by the Feynman rules.
In two dimensions GFT's  reduce to matrix models some of which \cite{GW,GW1} are ultraviolet 
complete \cite{GW2,GW3}. This opens up the tantalizing possibility that the GFT's themselves
are consistent and complete quantum field theories. 

The scenario of GFT as a fundamental quantum field theory recently received renewed 
attention. Partial power counting theorems and bounds \cite{FreiGurOriti,sefu1} have 
been obtained for the simplest GFT models. More accurate power counting 
theorems have been established \cite{sefu2, sefu3} for the 
``colored GFT's'' \cite{color, PolyColor} and recently
extended \cite{matteo}.

However there is a fundamental aspect of GFT's which has been little addressed so far
but has the potential to completely invalidate them: the topology of the gluings dual 
to GFT graphs. It hes been noted from some time \cite{DP-P} that GFT's generate
not only manifolds but also pseudo manifolds. As it is clear that space time
is a manifold, this is a rather unpleasant feature of GFT's. However
it is not critical: pseudo manifolds are related (one to one) to manifolds 
with boundary, and gravity makes perfect sense on the latter.
In fact in at least an approach to spin foams (and consequently GFT) such configurations are desirable, 
\cite{Alexander:2003kx}: the topological defects of pseudo manifolds can be interpreted 
as matter coupled to the gravitational background.

This paper addresses, in its full generality, the problem of the 
topology of gluings dual to GFT graphs.
An in depth study of this question will reveal a very serious issue which
has been largely ignored up to now: in all dimensions,
including three, there exist GFT graphs dual to
gluings which are {\it not} pseudo manifolds but correspond to much
more singular topologies. 
\begin{figure}[htb]
\centering{
\includegraphics[width=20mm]{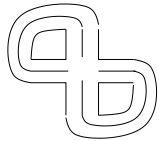}}
\caption{A graph {\it not} dual to a pseudo manifold.}
\label{fig:intro1}
\end{figure}
The simplest example of such a graph is presented in figure 
\ref{fig:intro1}. We will present in detail this example in section \ref{sec:wrap}, 
but for now it suffices to say that the Euler characteristic of its dual gluing is $-1$.
It is a fundamental result (which we recall in section \ref{sec:manifolds} for completeness) 
that the Euler characteristic of three dimensional pseudo manifolds is {\it allways} 
greater or equal to zero, thus this graph {\it can not} correspond to a pseudo manifold.

The pathological singularities we identify in this paper are {\it generic}, appearing
at arbitrary order in perturbations, and {\it dominate} in power counting.
This brings into question the usual GFT's status as ``fundamental'' quantum field theories: 
their effective behavior is dominated by pathological configurations.
In retrospect, when compared to these pathologies, the pseudo manifolds seem 
just a small nuisance one can live with.

However GFT's miss their target only by an inch. If one 
assumes that the gluing dual to a GFT graph is a 
{\it simplicial complex}\footnote{Or that it becomes 
one after subdivision.}, then all the pathologies disappear and the gluing 
is a pseudo manifold.

In order to salvage the GFT's as quantum field theories one must find some
way to eliminate the singular topologies. Restricting, by some condition at 
the level of the action the allowed gluings one can hope that only 
pseudo manifolds are created. However, finding a good restriction 
is a subtle question. For instance, requiring that the identification of 
two $(n-1)$ simplices respects orientations 
is largely insufficient: their $(n-2)$, $(n-3)$ etc. subsimplices are 
identified in a completely arbitrary way and generate pathologies.

A surprisingly simple solution to this problem is provided by the recently 
introduced ``colored'' GFT (CGFT) models . This model completely {\it eliminates} 
all the pathologies, yielding only pseudo manifolds in {\it any} dimension by an unique
prescription. Establishing this result is the ``raison d'\^etre'' of this paper.
\begin{figure}[htb]
\centering{
\includegraphics[width=40mm]{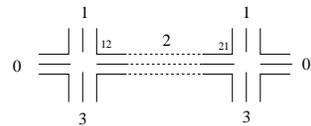}}
\caption{Colored GFT line.}
\label{fig:intro2}
\end{figure}
We will prove this in all the technical detail in section \ref{sec:newmodel}, 
but the profound reason which makes the colored models work is very intuitive. 
If we denote a strand (solid line in figure \ref{fig:intro2}) by the colors of 
the two halflines to which it belongs, the colored GFT lines will {\it allways} 
conserve the labels of the strands. In turn this will guarantee that
{\it all} subsimplices (of any dimension) are identified respecting their orientations. 
In retrospect the colored prescription is very natural: it is the simplest one which 
ensures this. We view this result as a very strong argument in support
of the colored GFT models.

This paper is organized as follows. In section \ref{sec:manifolds} we review some definitions and
classical results concerning normal simplicial pseudo manifolds. In section \ref{sec:clasic} we 
describe in detail the usual GFT's and their graphs, and introduce the link graphs in subsection 
\ref{sec:bal}. We detail at length the pathological wrapping singularities plaguing the usual GFT's 
in section \ref{sec:wrap}. In section \ref{sec:newmodel} we recall the colored GFT models and 
prove that they only generate graphs dual to normal simplicial pseudo manifolds. Finally, 
in section \ref{sec:conclusion}, we review the implications of our result.
We will only deal with closed GFT graphs, the generalization to open graphs \cite{PolyColor}
being immediate.

\section{Simplicial Pseudo Manifolds}
\label{sec:manifolds}

In this section we review some definitions and properties of normal simplicial pseudo manifolds
(following the notations of \cite{rus}) relevant to our subsequent analysis of GFT's.

A {\it finite abstract simplicial complex}\footnote{Or {\it simplicial complex}, for brevity.} is a finite set $A$ together with a 
collection $\Delta$ of subsets such that if $X\in \Delta$ and $Y\subseteq X$ then $Y\in \Delta$.

An element $v\in A$ such that $\{v\} \in \Delta$ is called a vertex of $\Delta$, and the set of all vertices
of $\Delta$ is denoted $V(\Delta)$. An element $\sigma\in \Delta$ is called a {\it simplex}. The proper
subsets $\tau$ of a simplex $\sigma$,  ($\tau\subset \sigma, \; \sigma\setminus \tau \neq \emptyset$) are 
called {\it faces} or {\it subsimplices} of $\sigma$.
Note that $\Delta$ is not a set but a collection (or a {\it multiset}), meaning that the same simplex can appear several times 
in $\Delta$. A subcomplex $\Delta'$ of $\Delta$ is a simplicial complex such that 
$\sigma\in\Delta'\Rightarrow \sigma \in \Delta$.
To any simplex in a simplicial complex one canonically associates several simplicial subcomplexes of
$\Delta$
\begin{itemize}
\item The {\it deletion} of $\tau$ is the abstract simplicial subcomplex of $\Delta$
\bea
 \text{dl}_{\Delta} (\tau) = \{\sigma \in \Delta | \; \tau \nsubseteq \sigma \} \; .
\eea 
\item The {\it link} of $\tau$ is the abstract simplicial subcomplex of $\Delta$
\bea
 \text{lk}_{\Delta} (\tau) = \{ \sigma \in \Delta | \; \sigma \cap \tau=\emptyset \text{ and } \sigma \cup \tau \in \Delta \} \; .
\eea 
\item The {\it closed star} of $\tau$ is the abstract simplicial subcomplex of $\Delta$
\bea
 \text{star}_{\Delta} (\tau) = \{\sigma\in \Delta | \;  \sigma \cup \tau \in \Delta \} \; .
\eea 
\end{itemize} 

The link and the closed star of a simplex $\tau$ are related by
\bea\label{eq:cap}
 \text{lk}_{\Delta}(\tau)=\text{star}_{\Delta}(\tau) \cap \text{dl}_{\Delta}(V(\tau)) \; \;,
\eea
as $\sigma \nsupseteq \{v\},\;  \forall \; v\in V(\tau)\Rightarrow \sigma\cap \tau =\emptyset$.

For any vertex $v$ of $\Delta$, and any simplex $\sigma \in \Delta$, either $\{v\} \nsubseteq \sigma$ 
or $\{v\} \cup \sigma = \sigma \in \Delta$, hence
\bea\label{eq:cup}
  \Delta=\text{star}_{\Delta}(v) \cup \text{dl}_{\Delta}(v) \;.
\eea

A simplex $\tau$ of a simplicial complex $\Delta$ has {\it dimension} $n$ (it is an $n$ simplex) if it has cardinality $n+1$. For instance, the vertices of $\Delta$ have dimension 0. We denote the number of simplices of dimension $p$ in $\Delta$ by $f^p(\Delta)$ (hence $f^{0}(\Delta)=|V(\Delta)|$) and its Euler characteristic by 
\bea
 \chi(\Delta) = \sum_{p \ge 0} (-1)^p f^p(\Delta) \; .
\eea
For any vertex $v$ eq. (\ref{eq:cap}) and (\ref{eq:cup}) imply that the Euler characteristic of a simplicial
complex respects 
\bea
 \chi(\Delta)&=&\chi\big{(}\text{star}_{\Delta}(v) \big{)} + \chi\big{(}\text{dl}_{\Delta}(v)\big{)}
 - \chi\big{(}\text{lk}_{\Delta}(v) \big{)} \nonumber\\
&=&1-\chi\big{(}\text{lk}_{\Delta}(v)\big{)} + \chi\big{(}\text{dl}_{\Delta}(v) \big{)}
\; , 
\eea
where in the last line we used $\chi\big{(}\text{star}_{\Delta}(v) \big{)}=1$ (see appendix \ref{app:prfs}) .

An {\it $n$-dimensional simplicial pseudo manifold} is a finite abstract simplicial complex with the following properties:
\begin{itemize}
\item it is {\it non-branching}: Each $(n-1)$ simplex is a face of precisely two $n$ simplices. 
\item it is {\it strongly connected}: Any two $n$ simplices can be joined by a ``strong chain''  
of $n$ simplices in which each pair of neighboring simplices have a common $(n-1)$ simplex. 
\item it is {\it pure} (it has {\it dimensional homogeneity}): Each simplex is a face of some $n$ simplex.
\end{itemize}

A pseudo manifold is called {\it normal} if all its links are pseudo manifolds. 
This condition can fail (see appendix \ref{app:prfs}) because the links
of a pseudo manifold, while always being pure, non branching simplicial complexes,
are not in general strongly connected. Crucial in the sequel is the following
property of three dimensional normal pseudo manifolds
\begin{proposition}\label{prop:euler}
 The Euler character of a three dimensional normal pseudo manifold $\Delta$ respects
    \bea
    \chi(\Delta) = |V(\Delta)|-\frac{1}{2}\sum_{i} \chi(\text{lk}_{\Delta}(v_i)) \;,
   \eea
  and $\chi(\text{lk}_{\Delta}(v_i)) \le 2$, thus $\chi(\Delta)\ge 0$.
\end{proposition}
{\bf Proof:}  Counting subsets of fixed cardinality shows that a pure simplicial complex respect
  \bea
   &&\sum_i f^2\bigl(\text{lk}_{\Delta}(v_i)\bigr) = 4 f^3(\Delta) \nonumber\\
   &&\sum_i f^1\bigl(\text{lk}_{\Delta}(v_i)\bigr) = 3 f^{2}(\Delta) \nonumber\\
   &&\sum_i f^0\bigl(\text{lk}_{\Delta}(v_i)\bigr) = 2 f^{1}(\Delta) \; .
  \eea

In a non branching simplicial complex in three dimensions every 3 simplex is bounded by four 2 simplices and every 2 simplex belongs to exactly two 3 simplices, hence
  \bea
   4 f^{3}(\Delta) = 2 f^{2}(\Delta) \; ,
  \eea
hence the Euler characteristic of a three dimensional pure, non branching simplicial complex  respects
   \bea
    \chi(\Delta) = f^{0}(\Delta)-\frac{1}{2}\sum_{i} \chi\bigl(\text{lk}_{\Delta}(v_i)\bigr) \; .
   \eea

If, furthermore, $\Delta$ is a normal pseudo manifold, all the links of its vertices are 
two dimensional pseudo manifolds. A link, $\text{lk}_{\Delta}(v_i)$, is strongly connected, 
hence there exists a ``strong tree'' of 1 simplices connecting all its 2 simplices. 
If one deletes the 1 simplices in the strong tree (and glues the 2 simplices into a patch),
the 0 simplices are still connected by (at least a tree of) the remaining 1 simplices, thus
\bea
   f^{1}\bigl(\text{lk}_{\Delta}(v_i)\bigr)  \ge 
\Big{[}f^2\bigl(\text{lk}_{\Delta}(v_i)\bigr)-1 \Big{]}+
\Big{[}f^0\bigl(\text{lk}_{\Delta}(v_i)\bigr) -1\Big{]} \; ,
\eea 
which achieves the proof.

\qed

\section{GFT Graphs}
\label{sec:clasic}

In this section we detail the Feynman graphs of the usual GFT models in $n$ 
dimensions and relate them with normal simplicial pseudo manifolds.

The usual $n$ dimensional GFT model is defined for a scalar 
field $\phi:G^{n}\rightarrow \mathbb{R}$, (with $G$ some Lie group), 
symmetric under permutations $\pi$ of its arguments 
and invariant under simultaneous left multiplication 
\bea
&& \phi(g_{\alpha^{}_{\pi(1)}},\dots, g_{\alpha^{}_{\pi(n)} } )= \phi(g_{\alpha^{}_1},\dots, g_{\alpha^{}_n})
   \; , \; \forall \pi \text{ permutation} \;, \nonumber\\
&& \phi(hg_{\alpha^{}_1},\dots, hg_{\alpha^{}_n})=\phi(g_{\alpha^{}_1},\dots, g_{\alpha^{}_n})\; ,
   \; \forall h\in G \; .
\eea

\begin{figure}[htb]
\centering{
\includegraphics[width=40mm]{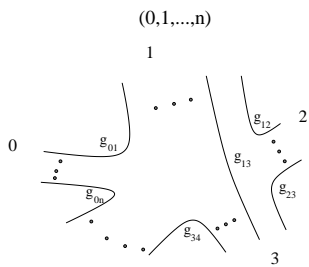}}
\caption{GFT vertex in $n$ dimensions.}
\label{fig:nDvertex}
\end{figure}

The GFT action in $n$ dimensions is \cite{DP-P}
\bea\label{eq:nD}
S&=&\frac{1}{2}\int [dg] \;
\phi_{\alpha^{}_0 \alpha^{}_1 \dots \alpha^{}_n} \;  \phi_{\alpha^{}_0\alpha^{}_1 \dots \alpha^{}_n} 
+S_{int} \; ,\nonumber\\
S_{int}&=& \frac{\lambda}{n+1}\int [dg] \; \phi_{\alpha^{}_{0n} \alpha^{}_{0n-1} \dots \alpha^{}_{01}} 
\dots  \phi_{\alpha^{}_{pp-1} \dots \alpha^{}_{p0} \alpha^{}_{pn} \dots \alpha_{pp+1}} \nonumber\\
&&\times \dots  \phi_{\alpha^{}_{nn-1} \dots \alpha^{}_{n0}} 
\eea 
where $\phi_{\alpha^{}_0\alpha^{}_1 \dots \alpha^{}_n} \equiv \phi(g_{\alpha^{}_0},g_{\alpha^{}_1},\dots, g_{\alpha^{}_n})$, 
and $g_{\alpha^{}_{ij}}=g_{\alpha^{}_{ji}}$ in $S_{int}$. The GFT vertex generated by $S_{int}$, is 
represented in figure \ref{fig:nDvertex}. 

Each field $\phi$ in $S_{int}$ is associated to a half line of the GFT vertex. Every two fields in $S_{int}$ 
share a group element, consequently every two half lines of the GFT vertex share a {\it strand} (depicted 
as a solid line in figure \ref{fig:nDvertex}). We label the half lines of the GFT vertex $0$, $1$ up to $n$ and
each strand by the (unordered) couple of labels of the two half lines which share it 
(that is the strand $ij$ is shared by the half lines $i$ and $j$). 

The GFT vertex $(0\dots n)$ is {\it dual} to an $n$ simplex $\{A_0,\dots A_n\}$. The half lines of the 
vertex {\it represent} the $(n-1)$ simplices bounding the $n$ simplex, namely the half line $i$
represents the simplex opposite to the vertex $A_i$\footnote{Throughout
this paper we denote by a hat the absence of a symbol in a list.}.
\bea
\{A_0,\dots A_n\} \setminus \{A_i \} \equiv \{A_0,\dots \widehat{A_i},\dots A_n \} \; .
\eea

The strand $ij$ represents the $(n-2)$ simplex {\it shared} by the two $(n-1)$ simplices 
$\{A_0,\dots \widehat{A_i},\dots A_n \} $ and $\{A_0,\dots \widehat{A_j},\dots A_n \}$,
that is
\bea
\{A_0,\dots \widehat{A_i},\dots, \widehat{A_j}, \dots A_n \} \; .
\eea 
In the sequel the GFT vertex will be called a {\it stranded vertex}, to emphasize its 
internal strand structure.

We use this opportunity to clarify a somewhat confusing point: the half lines
of GFT vertex are {\it not} dual to $(n-1)$ simplices, they are graphical 
representations. An $(n-1)$ simplex is dual to a GFT vertex in the appropriate
dimension. This distinction is crucial in order to understand the link 
graphs of section \ref{sec:bal}.

\begin{figure}[htb]
\centering{
\includegraphics[width=80mm]{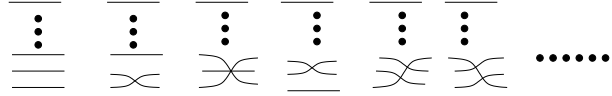}}
\caption{GFT lines in n dimensions.}
\label{fig:nDprop}
\end{figure}

The GFT propagator, generated by the quadratic part of the action (\ref{eq:nD}),
connects two GFT vertices via an arbitrary permutation of the strands. 
Some possible choices of GFT lines are presented in figure \ref{fig:nDprop}. 
The GFT lines represent the identification of two $(n-1)$ simplices and each 
permutation of the strands encodes one of the $n!$ possible ways to do this.
Like the GFT vertices, the GFT lines are {\it stranded} and have an internal 
structure.

To a GFT line $\ell_{v_A v_B}$ connecting two GFT vertices $v_A$ and $v_B$ (and
oriented from $v_A$ to $v_B$) dual to the $n$ simplices 
\bea 
\sigma^n_A=\{A_0,\dots, A_n \}\;,  \quad \sigma_B^n=\{B_0,\dots, B_n \} \; ,
\eea
we associate a function $\ell_{v_Av_B}:\{ 0,\dots,n\}\rightarrow \{0,\dots,n\}$
defined as follows. The line connects the half line $i$ of $v_A$ with some half line,
say $k$, of $v_B$. We set $\ell_{v_Av_B}(i)=k$. Also, the line connects the strand 
$ij$ of $v_A$ to some strand, say $kl$ of $v_B$. We set $\ell_{v_Av_B}(j)=l$. 
The function $\ell_{v_Av_B}$ encodes the identification of the two $(n-1)$ simplices
\bea
 \{A_0,\dots,\widehat{A_i},\dots  A_n \} ,  \{B_0,\dots,\widehat{B_{F(i)}},\dots  B_n \},
\eea
and {\it all their faces} via $ A_j = B_{F(j)}, \; \forall j\neq i \;$.

The perturbative development of GFT is indexed by 
{\it stranded} Feynman graphs $\cG$ generalizing the 
ribbon graphs of matrix models. A GFT graph $\cG$ is
dual to come gluing of $n$ simplices, denoted in the sequel $\Delta^{\cG}$.

The gluing $\Delta^{\cG}$ is a collection of $n$ simplices (and all their faces) 
modulo the identifications encoded in the lines. Clearly $X\in \Delta^{\cG}$ and
$Y\subset X$ then $Y\in \Delta^{\cG}$, thus $\Delta^{\cG}$ is very close to a
simplicial complex. However, in general, $\Delta^{\cG}$ is {\bf not} a simplicial complex.
Performing the identifications encoded in the lines one can end up identifying two a 
priori distinct vertices on the same $n$ simplex.  
Consequently, the elements $X\in \Delta^{\cG}$ are 
{\bf not} sets, but multisets\footnote{To add to the confusion recall that $\Delta$ itself {\it is} a multiset. Its  elements $X\in \Delta$, however, must be sets.}.

This is not always a problem. It is possible that, even if $\Delta^{\cG}$ is not a simplicial complex,
it is still topologically equivalent to some simplicial complex $\tilde{\Delta}^{\cG}$.
What is much less obvious is that sometimes $\Delta^{\cG}$ is {\it not} equivalent
to {\it any} simplicial complex.
This in turn leads to some very pathological singularities.

\subsection{Link graphs}\label{sec:bal}

The links defined for simplicial complexes generalize immediately to gluings.
The link of a $p$ simplex is a gluing of $(n-p-1)$ simplices,
hence it is dual to a GFT graph in $(n-p-1)$ dimensions. We call this  
graph a {\it link graph}. To construct it, consider the $p$ simplex $\sigma^{p}=\{A_{i_0},\dots A_{i_p},\}$
in a gluing. The contribution of the $n$ simplex
$\sigma^n=\{A_0,\dots A_n\}$ to its link consist of the simplex 
$\sigma^{n-p-1}=\sigma^{n} \setminus \sigma^p $ and all its faces.
The $n$ simplex $\sigma^n$ is dual to the GFT vertex $(0\dots n)$, therefore $\sigma^{n-p}$ is dual to 
the GFT vertex $(0\dots \widehat{i_0} \dots \widehat{i_p}\dots n)$ obtained 
by deleting all the half lines $i_0,\dots i_p$ {\it together with all their strands} in the 
initial GFT vertex $(0,\dots n)$. We call $(0\dots \widehat{i_0} \dots \widehat{i_p}\dots n)$
a descendant vertex of $(0\dots n)$. The $(n-p-1)$ link graphs are obtained by connecting 
the descendant $(n-p-1)$ vertices of all initial $n$ dimensional GFT vertices as 
dictated by the GFT lines.

\begin{figure}[htb]
\centering{
\includegraphics[width=80mm]{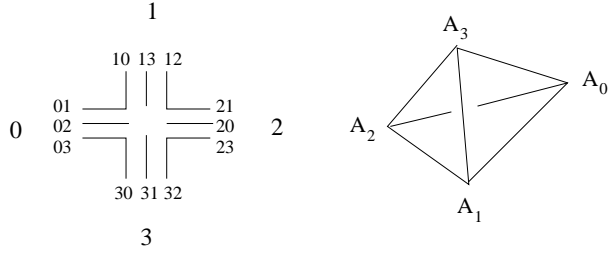}}
\caption{GFT vertex in 3 dimensions.}
\label{fig:3Dvertex}
\end{figure}

Consider the example of three dimensional GFT whose vertex and dual three simplex (tetrahedron) 
are presented in figure \ref{fig:3Dvertex}.

A vertex (say $A_0$) of the tetrahedron $\sigma^3=\{A_0,A_1,A_2,A_3\}$ is opposite to 
a triangle $\sigma^2= \sigma^3\setminus \{A_0\}= \{A_1,A_2,A_3\}$. This triangle is 
represented by a half line (the half line 0) in the GFT graph.
Two triangles (say $\{A_1,A_2,A_3\} $ and $ \{A_0,A_2,A_3\}$) share and edge on the tetrahedron 
(the edge $\{A_2,A_3\}$). This edge is represented by the strand common to the two half lines
(the strand $01$, common to the half lines $0$ and $1$).

\begin{figure}[htb]
\centering{
\includegraphics[width=70mm]{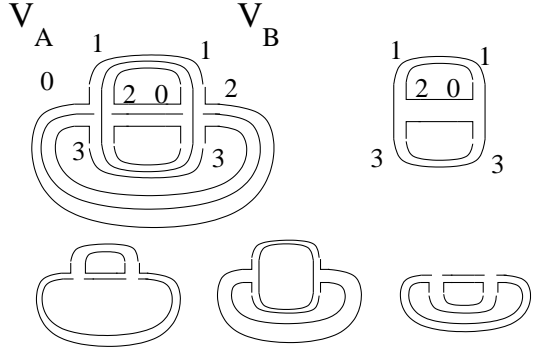}}
\caption{Link graphs in GFT.}
\label{fig:strandbub}
\end{figure}

Consider the example of the GFT graph in figure \ref{fig:strandbub}. Its dual gluing,
$\Delta^{\cG}$, consists of two tetrahedra $\{A_0,A_1,A_2,A_3\}$ and 
$\{B_{0},B_{1},B_{2},B_{3}\}$, and four identifications (hence four functions) associated to the lines
\bea
 \ell^{(1)}_{v_Av_B}(2)=0\;; && \ell_{v_Av_B}^{(1)}(1)=1,\; 
                 \ell_{v_Av_B}^{(1)}(0)=2, \ell_{v_Av_B}^{(1)}(3)=3 \; ,\nonumber\\
 \ell_{v_Av_B}^{(2)}(1)=1\;; && \ell_{v_Av_B}^{(2)}(2)=0,\; 
\ell_{v_Av_B}^{(2)}(3)=3, \ell_{v_Av_B}^{(2)}(0)=2 \; ,\nonumber\\
 \ell_{v_Av_B}^{(3)}(0)=2\;; && \ell_{v_Av_B}^{(3)}(1)=1,\; 
  \ell_{v_Av_B}^{(3)}(2)=0, \ell_{v_Av_B}^{(3)}(3)=3 \; ,\nonumber\\
 \ell_{v_Av_B}^{(4)}(3)=3\;; && \ell_{v_Av_B}^{(4)}(0)=2,\; \ell_{v_Av_B}^{(4)}(1)=1, 
\ell_{v_Av_B}^{(4)}(2)=0 \; ,
\eea
where the first column is the half line of $v_A$ from which each line originates, and the 
subsequent columns indicate the various identifications of $A$'s with $B$'s.
The line $\ell^{(1)}_{v_Av_B}$ for example encodes he identifications
\bea
&& \{A_1,A_0,A_3\}=\{B_1,B_2,B_3\} \quad \{A_1,A_0 \}=\{B_1,B_2 \} \nonumber\\
&& \{ A_1,A_3 \}=\{B_1,B_3 \} \quad \{A_0,A_3\}=\{ B_2,B_3\} \nonumber\\ 
&&\{A_1 \} = \{B_1\} \quad \{A_0 \}=\{B_2\} \quad \{A_3\}=\{B_3\}    \; .
\eea
The reader can convince himself that, after performing also the identifications
corresponding to $\ell^{(2)}_{v_Av_B}$, $\ell^{(3)}_{v_Av_B}$ and $\ell^{(4)}_{v_Av_B}$,
the gluing writes
\bea
 \Delta^{\cG}=&\Big{\{}&\emptyset, \{A_0\},\{A_1 \}, \{A_2\}, \{A_3\}, \{A_0,A_1\}, \{A_0,A_2\}, \{A_0,A_3\}, 
 \nonumber\\
&& \{A_1,A_2\}, \{A_1,A_3\}, \{A_2,A_3\}, \{A_0,A_1,A_2\},  \nonumber\\
&& \{A_0,A_1,A_3\}, \{A_0,A_2,A_3\}, \{A_1,A_2,A_3\},
\nonumber\\
&& \{A_0,A_1,A_2,A_3\},  \{A_0,A_1,A_2,A_3\} \Big{\}} \; .
\eea
Note that $\Delta^{\cG}$ is a multiset (the two $3$ simplices have exactly the same vertices) 
and one can check that this gluing is a simplicial complex. The link of $A_0$, for instance writes
\bea
 \text{lk}_{\Delta^{\cG}}(A_0)=&\Big{\{}&\emptyset,\{A_1 \}, \{A_2\} \{A_3\}, \{A_1,A_2\}, 
 \{A_1,A_3\}, \{A_2,A_3\},
\nonumber\\
&& \{A_1,A_2,A_3\},  \{A_1,A_2,A_3\} \Big{\}} \; .
\eea

It is in fact easier to access directly the link graph dual to 
$\text{lk}_{\Delta^{\cG}}(A_0)$ starting from $\cG$. 
To build the link graph dual to $\text{lk}_{\Delta}(A_0)$ we distinguish the labels on the vertices 
$\cV_A$ and $\cV_B$ by a lower index. Take the descendant vertex $1_A2_A3_A$ obtained by deleting the 
half line $0_A$ (and all its strands) of $(0_A1_A2_A3_A)$. The half lines of the descendant vertex
inherit the labels of the corresponding GFT half lines, $1_A$, $2_A$ and $3_A$, and a pair of 
descendant half lines share a strand ($1_A$ and $2_A$ share the strand $1_A2_A$, etc.).

The half line $1_A$ of the descendant vertex $1_A2_A3_A$ connects to the half line $1_B$ of the 
descendant vertex $1_B0_B3_B$ (obtained by deleting $2_B$) of $0_B1_B2_B3_B$. 
Similarly, the half line $2_A$ connects to $0_B$ and $3_A$ to $3_B$ of the same
descendant vertex $1_B0_B3_B$. The two descendent's vertices thus form a connected 
graph, dual to the link $\text{lk}_{\Delta}(A_0)$.

By construction, every GFT vertex in three dimensions has four descendants vertices in the link graphs, 
thus the dual gluing of any GFT graph respects 
\bea\label{eq:bal1}
\sum_i f^2\bigl(\text{lk}_{\Delta}(v_i)\bigr) = 4 f^3(\Delta)\; .
\eea 
Also, any GFT line always has three descendants (any two strands of a GFT line yield
a descendant line in some link graph), hence the dual gluing respects
\bea\label{eq:bal2}
\sum_i f^1\bigl(\text{lk}_{\Delta}(v_i)\bigr) = 3 f^2(\Delta)\;.
\eea 
Moreover each strand on the GFT vertex has two descendant strands in the link graphs.

The main result of this section is synthesized in the flowing lemma.

\begin{lemma}\label{lem:simpmanif}
If a gluing $\Delta^{\cG}$ dual to a $n$ dimensional connected GFT graph $\cG$ is a simplicial complex 
then it is a normal pseudo manifold.
\end{lemma}

Let us comment on this lemma before proving it. In the mathematical 
literature there are numerous results concerning pseudo manifolds (notoriously, for example,
in three dimensions they only present isolated singularities). Whereas this
results hold for some graphs, they {\it fail} in general. For instance, 
the Betti numbers and boundary operators, relevant for power counting 
estimates, can be defined only for graphs dual to pseudo manifolds.
They {\it make no sense} for arbitrary GFT graphs. 

\medskip

\noindent{\bf Proof of lemma \ref{lem:simpmanif}:}
A gluing dual to a connected GFT graph is always pure and strongly connected.
The GFT lines either connect two different GFT vertices or are tadpole lines 
(they start and end on the same GFT vertex). Thus in the dual gluing the
$(n-1)$ simplices either separate two distinct $n$ simplices or belong twice to the
same $n$ simplex.

If a $(n-1)$ simplex belongs twice to the same $n$ simplex, then in its corresponding 
gluing at least two a priori distinct vertices of the $n$ simplex are identified. 
Thus the $n$ simplex is not represented by a set in $\Delta^{\cG}$, but my a multiset 
and $\Delta^{\cG}$ is not a simplicial complex. Consequently, if $\Delta^{\cG}$ is a 
simplicial complex, then all its $(n-1)$ simplices bound exactly two $n$ simplices, 
therefore $\Delta^{\cG}$ is non branching thus a 
simplicial pseudo manifold.

The link graphs of a GFT graph $\cG$ are also GFT graphs (of lower dimensions). If $\cG$ has no tadpole lines, 
none of its links can have tadpole lines (the lines of link graphs are descendants of lines of $\cG$). 
The same reasoning as before holds for all the link graphs. Thus all the links of
$\Delta^{\cG}$ are also pseudo manifolds. Therefore $\Delta^{\cG}$ is a normal 
pseudo manifold.

\qed

As a last remark, note that we used the fact that if a GFT graph 
is a simplicial complex then it has no tadpole line.
If, however, a graph has no tadpole lines, its dual gluing might still 
not be simplicial complex: two vertices on a $n$ simplex could be identified after a 
longer sequence of gluings of lines (see section \ref{sec:wrap} for examples). 

\section{Wrapping Singularities in GFT Graphs}\label{sec:wrap}

We will detail the singularities of GFT graphs in three dimensions. 
We will present several examples of three dimensional GFT graphs whose dual gluings 
do not respect proposition \ref{prop:euler}, namely
 \bea
    \chi(\Delta) \neq |V(\Delta)|-\frac{1}{2}\sum_{i} \chi(\text{lk}_{\Delta}(v_i)) \;.
 \eea
As the Euler characteristic is a 
topological invariant these gluings are {\it not} homeomorphic 
to pseudo manifolds. We will prove that whenever a GFT graph presents a
certain type of singularity (we baptize wrapping singularity) it will
not respect proposition \ref{prop:euler}. We will show that these singularities are generic 
(they appear at arbitrary high order in perturbations). 
These problems reappears in all higher dimensions, as similar 
singularities in the link graphs prevent any higher dimensional gluing 
from being a normal pseudo manifold.

\begin{figure}[htb]
\centering{
\includegraphics[width=60mm]{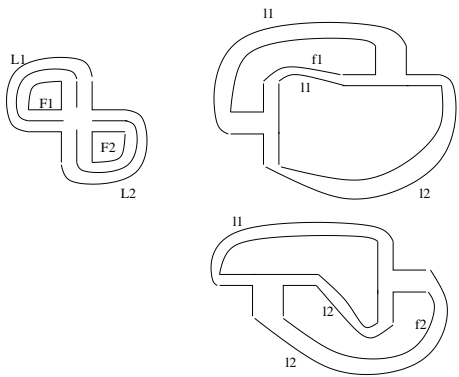}}
\caption{A GFT graph $\cG^1$ dual to a gluing with negative Euler characteristic.}
\label{fig:singgraph}
\end{figure}

Consider the GFT graph $\cG^1$ represented in figure \ref{fig:singgraph}.
The lines applications are
\bea
  l_{v_Av_A}^{(1)}(0)=1, \qquad  l_{v_Av_A}^{(1)}(1)=0,\; l_{v_Av_A}^{(1)}(2)=3,\;
 l_{v_Av_A}^{(1)}(3)=2 \; ,\nonumber\\
  l_{v_Av_A}^{(2)}(2)=3, \qquad  l_{v_Av_A}^{(2)}(0)=1,\; l_{v_Av_A}^{(2)}(1)=0,\; l_{v_Av_A}^{(2)}(3)=2 \; ,
\eea
where, again, the first column presents the half lines identified by the lines $l_{v_Av_A}^{(1)}$ and 
$l_{v_Av_A}^{(2)}$. Denoting $A_0=A_1=\alpha, A_2=A_3=\beta$, the dual gluing writes
\bea\label{eq:cg1}
 \Delta^{\cG^1} &=& 
\Big{\{} \{\alpha,\alpha,\beta, \beta \}, \{ \alpha,\alpha,\beta \} ,\{\alpha,\beta,\beta\},
\nonumber\\
&& \{\alpha,\alpha\},\{\alpha,\beta\},\{\alpha,\beta \}, \{\beta,\beta\},
\{\alpha\} , \{\beta\},\emptyset \Big{\}}.
\eea
Note that the 3 simplex of this gluing is not a set, hence $\Delta^{\cG^1}$ is not a simplicial complex. The
Euler characteristic of $\Delta^{\cG^1}$ is $\chi(\Delta^{\cG^1})=-1<0$ which breaks 
proposition \ref{prop:euler}. Therefore $\Delta^{\cG^1}$ is a first example of a gluing
{\it not} homeomorphic to a pseudo manifold. 

\begin{figure}[htb]
\centering{
\includegraphics[width=40mm]{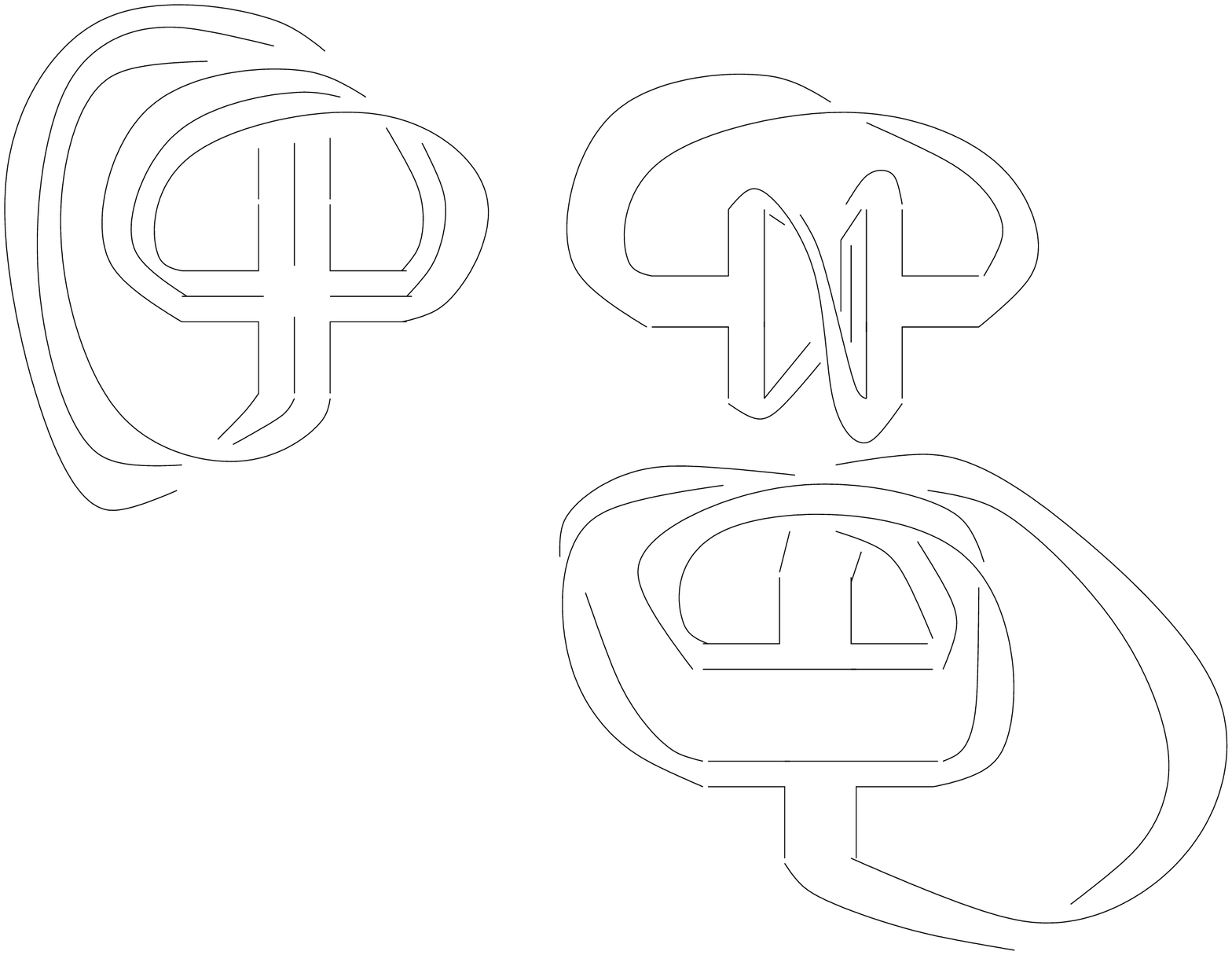}}
\caption{A second singular graph $\cG^2$.}
\label{fig:singgraph1}
\end{figure}

Proposition \ref{prop:euler} fails again for the graph $\cG^2$ of figure \ref{fig:singgraph1}, which
is in fact related by symmetry to $\cG^1$. The Feynman amplitude of these graphs is
\bea
 A_{\cG^1} = A_{\cG^2} =[\delta^{\Lambda}(e)]^2 \; ,
\eea 
where $\delta^{\Lambda}$ is a suitable cutoffed delta function on the group $G$, and $e$ is the identity 
element of $G$ (see \cite{FreiGurOriti,sefu1} for details on the computation of 
Feynman amplitudes in GFT).

\begin{figure}[htb]
\centering{
\includegraphics[width=40mm]{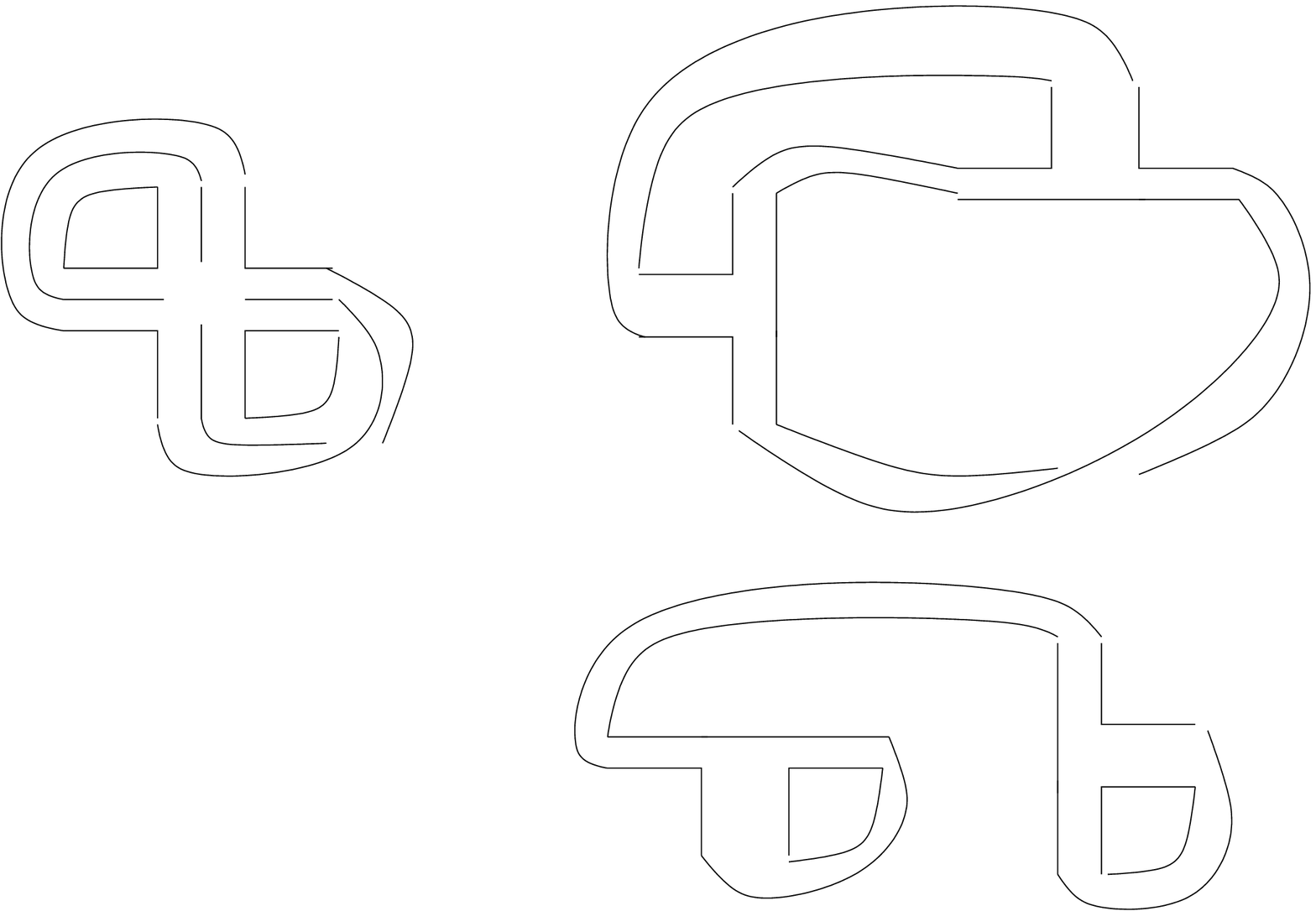}}
\caption{A third singular graph $\cG^3$.}
\label{fig:singgraph2}
\end{figure}

The graph $\cG^3$ in figure \ref{fig:singgraph2} has a planar link ( with Euler characteristic is $2$) 
and a non orientable one (with Euler characteristic $1$), hence $V-\frac{1}{2}\sum_{i}\chi(\text{lk}v_i)=1/2$ 
which is not even an integer. Its amplitude is
\bea
 A_{\cG^{3}}=\delta^{\Lambda}(e) \; .
\eea

\begin{figure}[htb]
\centering{
\includegraphics[width=40mm]{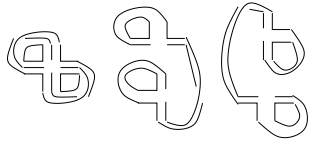}}
\caption{A graph $\cG^4$ dual to a sphere $S^3$.}
\label{fig:reggraph}
\end{figure}

At first order one also has a GFT graph dual to a gluing homeomorphic to a pseudo manifold (in fact homeomorphic to 
the manifold $S^3$), presented in figure \ref{fig:reggraph}.

The reader should not be distracted by the twists of the lines in the link graphs: they can be undone by flipping 
either of the end vertices. We prefer to represent the twists explicitly so that the reader can easily 
identify the the descendant vertices in the link graphs. In detail the line applications are
\bea
 &&\ell^{(1)}(0) = 1 \qquad \ell^{(1)}(3)=3,\;\ell^{(1)}(2)=2,\;\ell^{(1)}(1)=0, \nonumber\\
 &&\ell^{(2)}(2) = 3 \qquad \ell^{(2)}(1)=1,\;\ell^{(2)}(0)=0,\;\ell^{(2)}(3)=2 \; . 
\eea
The gluing dual to this graph writes (denoting $A_0=A_1=\alpha$ and $A_2=A_3=\beta$) 
\bea
 \Delta^{\cG^3}&=& 
\Big{\{} \{\alpha,\alpha,\beta, \beta \}, \{ \alpha,\alpha,\beta \} ,\{\alpha,\beta,\beta\},
\nonumber\\
&& \{\alpha,\alpha\},\{\alpha,\beta\}, \{\beta,\beta\},
\{\alpha\} , \{\beta\},\emptyset \Big{\}}. \; ,
\eea
to be compared with eq. (\ref{eq:cg1}). This amplitude of $\cG^4$ is
\bea
 A_{\cG^4}=\delta^{\Lambda}(e) \; .
\eea 

The analysis of these first four examples of graphs leads to the flowing conclusions: 
\begin{itemize}
 \item At first order, graphs not dual to pseudo manifolds are {\it larger} in power 
counting than graphs dual to pseudo manifolds. At arbitrary order, a graph obtained by star subdivisions 
(``one-four moves'') of $\cG^1$ will consistently have one extra power of 
$\delta_{\Lambda}(e)$ with respect to the similar graph obtained from $\cG^2$. 
 \item Restricting the permutations of strands allowed on the three dimensional 
GFT lines {\it does not} solve the problem: there exist singular graphs generated by even as well as odd permutations of the strands. Although (as we will see in the sequel) this idea is part of the solution, 
by itself it is insufficient. 
\end{itemize}

\begin{figure}[htb]
\centering{
\includegraphics[width=50mm]{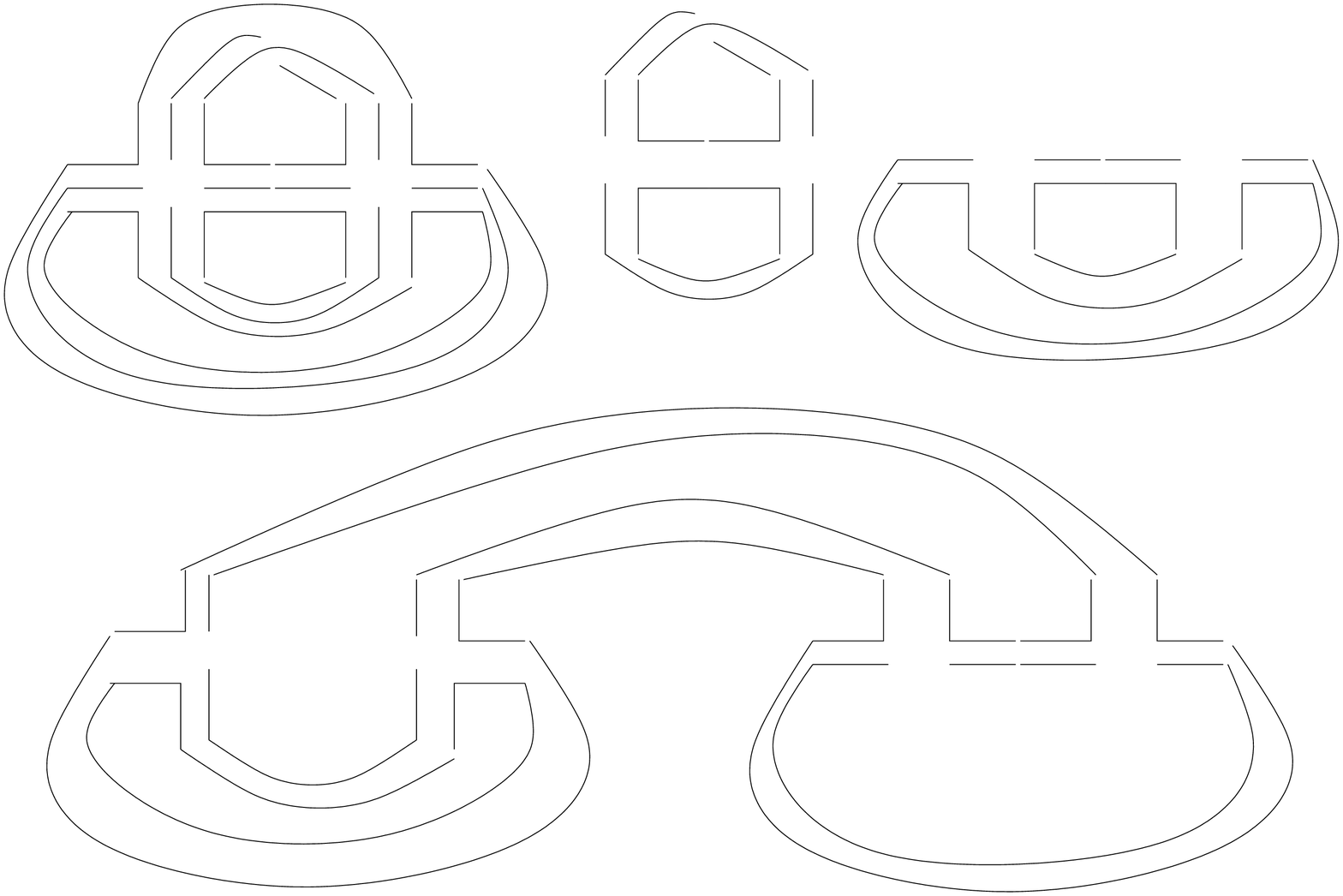}}
\caption{A singular graph $\cG^5$ with no tadpole lines.}
\label{fig:singgraph3}
\end{figure}

As all the examples we presented so far exhibit tadpole lines, the reader might still hope that the 
singularities are just an artifact of these tadpoles. This is not true, the example of \ref{fig:singgraph3}
presents a graph with no tadpole lines, whose links have Euler characteristics 1,2 and 2, hence
$V-\frac{1}{2}\sum_{i}\chi(\text{lk}v_i)=\frac{1}{2}$ again.

The proposition \ref{prop:euler} fails for arbitrary GFT graphs because the detailed balance 
crucial for its proof 
\begin{itemize}
 \item each GFT vertex has four descendant vertices in the link graphs.
 \item each GFT line has three descendant lines in the link graphs.
 \item each GFT face has two descendant faces in the link graphs.
\end{itemize}
{\it does not hold} in general. The attentive reader will recall the correct balance 
encoded in eq. (\ref{eq:bal1}) and  (\ref{eq:bal2})
\begin{itemize}
\item each GFT vertex has four descendant vertices in the link graphs.
 \item each GFT line has three descendant lines in the link graphs.
 \item each {\it strand} on a GFT vertex has two descendant {\it strands} in the link graphs.
\end{itemize}

The faces are {\it closed strands}, and in all the singular cases we presented the two descendant of 
some strand on a GFT vertex belong to {\it only one} face in the link graph. For instance in figure 
\ref{fig:singgraph} we denoted $F_1$ and $F_2$ two GFT faces, and $f_1$ and $f_2$ their 
{\it unique} descendant faces in the link graphs. The faces of the link graphs ($f_1$ and $f_2$) wrap twice around 
the GFT faces ($F_1$ and $F_2$), hence the name ``wrapping singularities''. The reader can check that this phenomenon is present in all the examples we presented.

\begin{figure}[htb]
\centering{
\includegraphics[width=40mm]{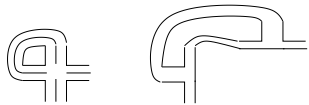}}
\caption{A subgraph leading to a wrapping singularity.}
\label{fig:singgraphfin}
\end{figure}

\begin{figure}[htb]
\centering{
\includegraphics[width=40mm]{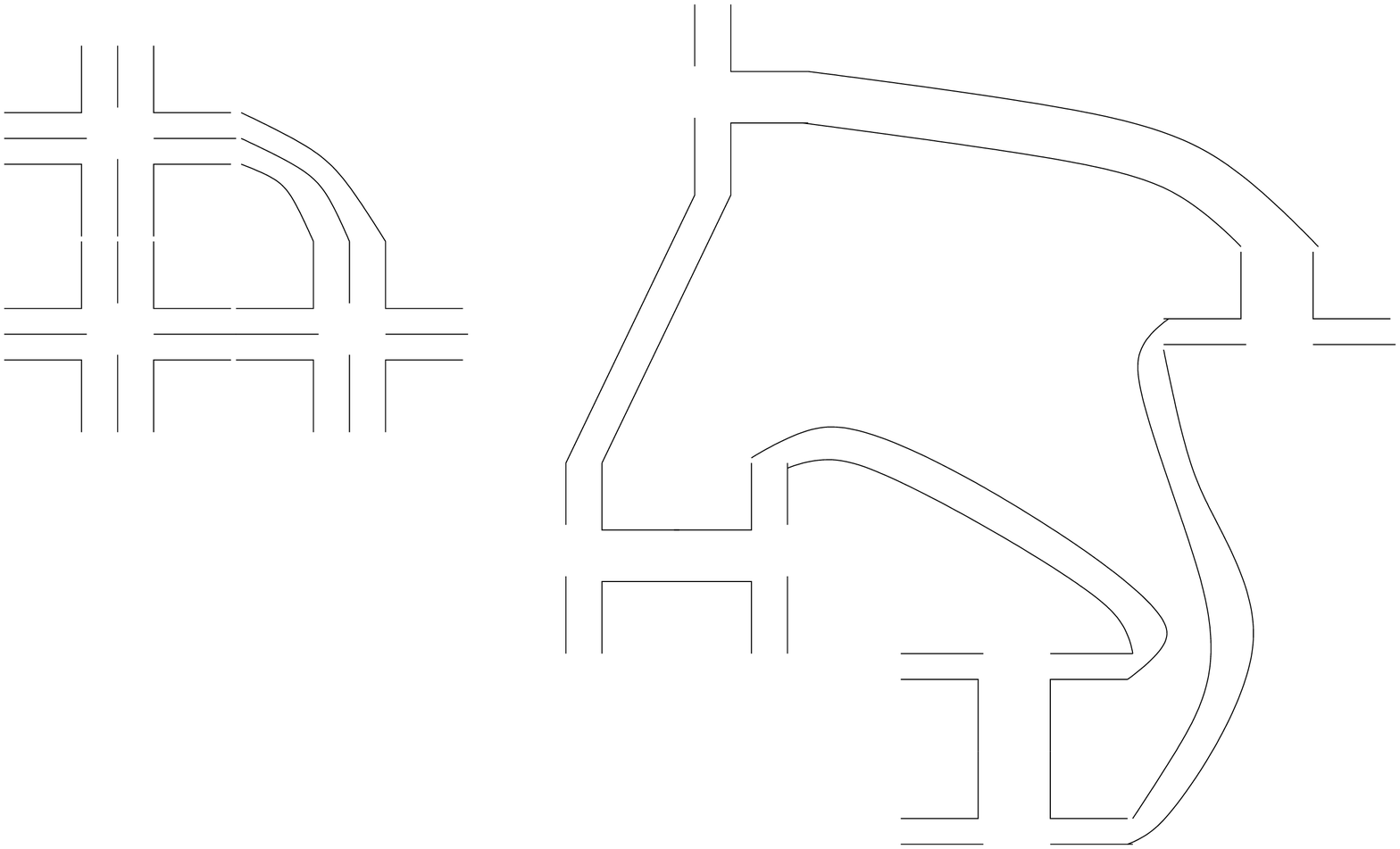}}
\caption{A second subgraph leading to a wrapping singularity.}
\label{fig:singgraphfinal}
\end{figure}

Whenever such singularities are present $\sum_i f^0\bigl(\text{lk}_{\Delta^{\cG}}(v_i)\bigr) < 2 f^{1}(\Delta^{\cG})$ and
$\Delta^{\cG}$ does {\it not} respect proposition \ref{prop:euler} hence it is 
{\it not} homeomorphic to a pseudo manifold. The wrapping singularities are generic in 
GFT: a graph having subgraphs like the ones  in figure \ref{fig:singgraphfin}, \ref{fig:singgraphfinal} 
will have a wrapping singularity. 

\begin{figure}[htb]
\centering{
\includegraphics[width=60mm]{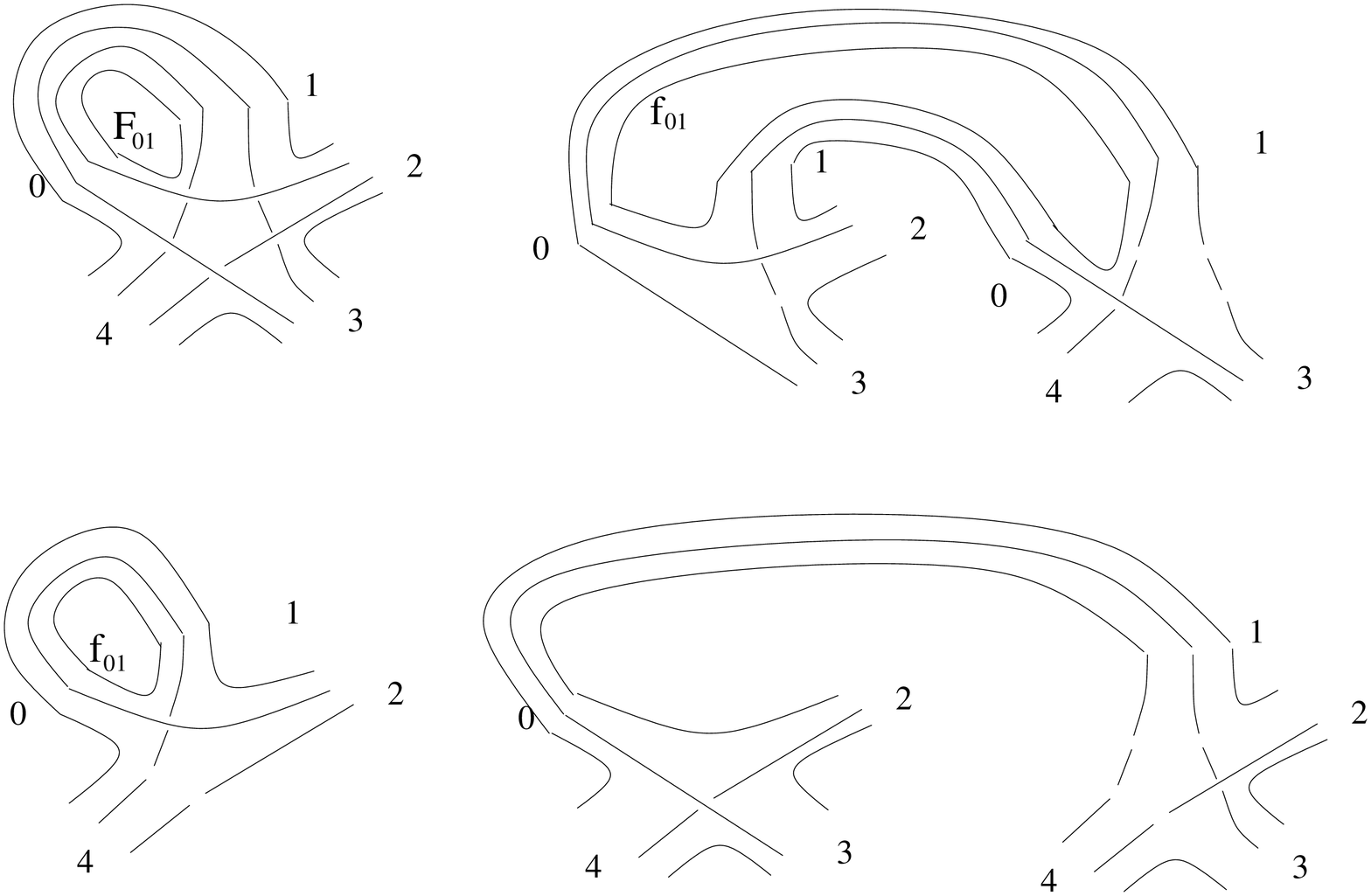}}
\caption{A four dimensional graph having a wrapping singularity.}
\label{fig:4D}
\end{figure}

In figure \ref{fig:4D} we give an example of a four dimensional graph having a wrapping singularity. The face $F_{01}$ has only two three dimensional descendants (instead of three), denoted both $f_{01}$, and one of them 
wraps twice around $F_{01}$.

The situation looks bleak for GFT's. The graphs with wrapping singularities are 
{\it large} in power counting and {\it generic}.
Singular graphs dominate the ``low energy'' effective behavior of GFT's and render them 
unsatisfactory.

\section{The Colored GFT Graphs and Pseudo Ma-nifolds}\label{sec:newmodel}

In this section we prove that colored GFT (CGFT) model \cite{color} {\it completely} 
solves the problem of 
non pseudo manifold graphs in a single stroke, in {\it arbitrary} dimension. By coloring
our quantum field we introduce a combinatorial constraint in all its graphs and completely
eliminate the wrapping singularities. Moreover, once the combinatorial constraints are
properly understood the proof that all CGFT graphs are dual to normal pseudo manifolds is practically
tautological. For this reason the CGFT model is, in our opinion, {\it the appropriate} GFT model 
one should always consider when treating GFT's as quantum field theories.

In $n$ dimensions, the colored GFT model is defined by $n+1$ pairs of fermionic 
(or complex bosonic) fields $\psi^p, \bar{\psi}^p:G^n\rightarrow \mathbb{G} \text{ or }\mathbb{C}$, invariant
under left group multiplication of the argument, and with {\it no} symmetry properties. The action
of the colored GFT writes
\bea\label{eq:color}
 S&=&\frac{1}{2}\int [dg] \;
\sum_{p=0}^{n}\psi^{p}_{\alpha^{}_0 \alpha^{}_1 \dots \alpha^{}_n} \;  
 \bar{\psi}^{p}_{\alpha^{}_0\alpha^{}_1 \dots \alpha^{}_n} 
+S_{int} +\bar{S}_{int} \; ,\nonumber\\
S_{int}&=& \lambda \int [dg] \; \psi^0_{\alpha^{}_{0n} \alpha^{}_{0n-1} \dots \alpha^{}_{01}} 
\dots  \psi^p_{\alpha^{}_{pp-1} \dots \alpha^{}_{p0} \alpha^{}_{pn} \dots \alpha_{pp+1}} \nonumber\\
&&\times \dots  \psi^n_{\alpha^{}_{nn-1} \dots \alpha^{}_{n0}} \; , 
\eea
and $\bar{S}_{int}$ has the same form as $S_{int}$ with $\psi$ replaced by $\bar{\psi}$. The index $p$ on 
each field is a {\it color} index and we denote the set of all colors 
$\cC^{n+1}=\{0,\dots,n\}$.

The interaction part of the colored GFT model has two terms and generates two vertices:
the positive vertex, involving only $\psi$'s, represented in figure \ref{fig:nDvertex} (where
the labels $0,\dots, n$ become now colors), and the negative vertex, involving only 
$\bar{\psi}$, with colors turning anticlockwise around it. 
The propagator of the model has $n$ {\it parallel} strands and always connects two 
half lines of the same color, one on a positive and one on a negative vertex. 
We orient all lines from positive to negative vertices. 

\begin{figure}[htb]
\centering{
\includegraphics[width=60mm]{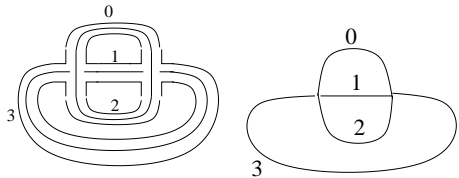}}
\caption{The unique second order CGFT graph $\cG$.}
\label{fig:exemplu}
\end{figure}

The strand structure of the vertex and propagator is rigid thus a CGFT graph
admits a simplified representation as a {\it colored} graph. The colored graph is
obtained by collapsing all the strands of the lines in ``thin'' lines, and all the strands of the
vertices in point vertices. Conversely, given a colored graph with thin lines and point vertices one 
can reconstruct the stranded graph associated to it. Figure \ref{fig:exemplu} depicts a CGFT 
graph either as a stranded graph (on the left) or as colored graph (on the right). 

A CGFT graph $\cG$ comes equipped with a natural family of subgraphs, called the 
{\it $p$-bubbles}. A $p$-bubble is a connected subgraph of $\cG$ made {\it only} of
lines of colors in $\cC^p$ for some subset $\cC^p \subset \cC^{n+1}$ of cardinality $|\cC^p|=p$.
We denote a $p$-bubble with colors $\cC^p$ and vertices $\cV$ by $\cB^{\cC^p}_{\cV}$.

\begin{figure}[htb]
\centering{
\includegraphics[width=100mm]{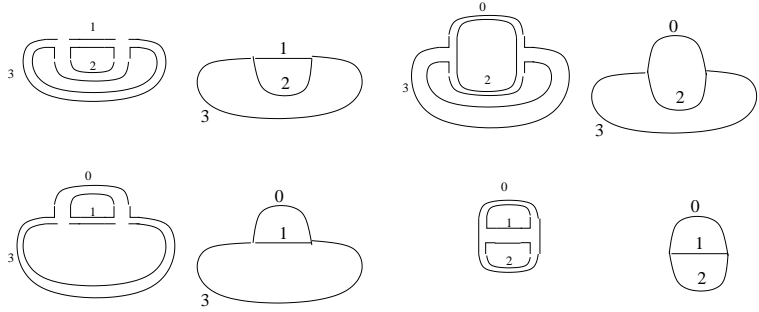}}
\caption{The 3-bubbles of $\cG$ represented as stranded or as colored graphs.}
\label{fig:exemplubub}
\end{figure}

Clearly the $0$-bubbles of a graph are its vertices and the $1$-bubbles are its lines. For
$p\ge 2$ the $p$-bubbles admit two graphical representations, either as colored graphs
or as stranded graphs. In the stranded graph representation one only draws the strands 
common to the lines of colors $\cC^p$. The colored and stranded representation of the 
$3$-bubbles of the graph $\cG$ in figure \ref{fig:exemplu} are depicted in figure 
\ref{fig:exemplubub}.

The $p$-bubbles are themselves colored GFT graphs in $(p-1)$ dimensions. Comparing
figure \ref{fig:exemplubub} with figure \ref{fig:strandbub}, we note that for this graph
the $3$-bubbles correspond to the link graphs. This is in fact a general result for $p\ge 2$

\begin{figure}[htb]
\centering{
\includegraphics[width=80mm]{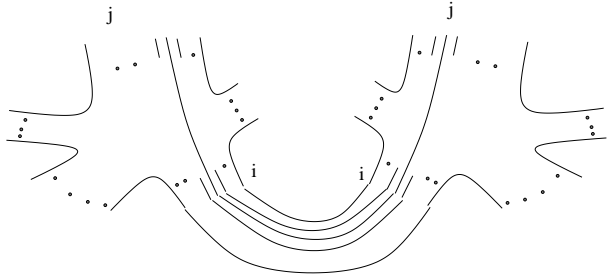}}
\caption{A colored GFT line.}
\label{fig:colorstrand}
\end{figure}

\begin{theorem}
  For $p\ge 2$, the $p$-bubbles of a CGFT graph are the link graphs of the $(n-p)$ simplices in the gluing $\Delta^{\cG}$. 
\end{theorem}
\noindent{\bf Proof:} Consider two vertices $v_A$ (positive) and $v_B$ (negative) connected 
by a line of color $i$ (see figure \ref{fig:colorstrand}) in an $n$-dimensional GFT graph $\cG$. 

This drawing essentially proves the result.  As the line has only parallel strands and connects opposite vertices,
the strand $ji$, common to the half lines $j$ and $i$ on $v_A$ necessarily connects with the strand $ij$ on $v_B$.
This holds for all lines, therefore the labels $ij$ are {\it conserved} all along the strand. 
This is the fundamental difference between the usual GFT graphs and the CGFT graphs and render 
the latter much better behaved.

The vertex $v_A$ is dual to some simplex $\sigma^n=\{A_0,\dots A_n\}$. Consider one of its $(n-p)$ subsimplices 
\bea
\sigma^{n-p}=\{A_0,\dots, A_n\}\setminus \{ A_{i_{1}},\dots, A_{i_p}\} \; .
\eea
Following subsection \ref{sec:bal}, the contribution of $v_A$ to the
link graph of $\sigma^p$ is the $p$ dimensional GFT vertex (descendant of $v_A$) with labels 
$(i_1 \dots i_p )$. But, as the colors of strands are conserved, this vertex will 
{\it always} connect {\it only} with the link vertex $( i_1 \dots i_p)$ 
descending from $v_B$. 

The link graphs are exactly the connected $p$ dimensional 
GFT graphs formed by lines and strands with colors  
$\{i_1,  \dots i_p \}$, hence the $p$ bubbles of $\cG$.

\qed

We are now in the position to state and prove the core result of this paper.

\begin{theorem}\label{th:pseudo}
  Any connected $n$ dimensional CGFT graph is dual to a normal simplicial pseudo manifold.
\end{theorem}
\noindent{\bf Proof} By lemma \ref{lem:simpmanif} it is enough to prove that the gluing
$\Delta^{\cG}$ dual to any CGFT graph is a simplicial complex. This is trivial once the
appropriate notations are introduced.

The 0 simplices (vertices) of the gluing $\Delta^{\cG}$ are dual to 
the $n$-bubbles of the CGFT graph, 
$V(\Delta^{\cG}) = \{\cB^{\cC^{n+1}\setminus\{p\}}_{\cV} \} $.

The 1 simplices of $\Delta^{\cG}$ the $(n-1)$-bubbles of $\cG$. Consider one of 
the $(n-1)$ bubbles of $\cG$, say $\cB^{\cC^{n+1}\setminus\{p,q\}}_{\cV}$. This bubble
is subgraph of $\cG$, hence there exists an unique subgraph obtained by adding all the lines 
of color $p$ for incident on the vertices $\cV$ 
and then closing the entire connected component with colors $\cC^{n+1}\setminus \{q\}$, that is
\bea
 \forall \; \cB^{\cC^{n+1}\setminus\{p,q\}}_{\cV}, \; \exists \;! \;
\cB^{\cC^{n+1}\setminus\{q\}}_{\cV^{\widehat{q}}}, \; \cV\subset \cV^{\widehat{q}} \; ,
\eea
where we denoted $\cV^{\widehat{q}}$ the (unique) maximal set of vertices connected by lines of 
colors $\cC^{n+1}\setminus \{q\}$ and containing $\cV$. The same holds for the lines of color $p$. 
Pick any vertex $v_A\in \cV$, dual to a $n$ simplex $\{A_0,\dots A_n\}$. 
The $1$ simplex dual to $\cB^{\cC^{n+1}\setminus\{p,q\}}_{\cV}$ is
$\{A_p,A_q\}$, and the $0$ simplex dual to $\cB^{\cC^{n+1}\setminus\{q\}}_{\cV^{\widehat{q}}}$
is $\{A_q\}$. Therefore the 1 simplex dual to $\cB^{\cC^{n+1}\setminus\{p,q\}}_{\cV}$ writes
\bea
 \Big{\{} \cB^{\cC^{n+1}\setminus\{q\}}_{\cV^{\widehat{q}}}, \cB^{\cC^{n+1}\setminus\{p\}}_{\cV^{\widehat{p}}} \Big{\}}, \quad
\cV\subset \cV^{\widehat{q}} \cap \cV^{\widehat{p}} \; .
\eea

Similarly, a $p$ simplex $\sigma^p$ is dual to a $(n-p)$ bubble 
$ \cB^{\cC^n\setminus \{i_0,\dots i_p \}}_{\cV} $, and for each $(n-p)$ bubble
and each color $i_q$, there exists an unique subgraph with $n$ colors obtained 
by adding the lines of colors all colors {\it except} $i_q$ incident on 
the vertices $\cV$ and completing the connected component with lines of all colors
{\it except} $i_q$ (whose set of vertices we denote $\cV^{\widehat{i_q}}$)
\bea
\cB^{\cC^{n+1}\setminus\{ i_q \}}_{\cV^{ \widehat{i_q}}},
\quad \cV\subset \cV^{\widehat{i_q}}\; .
\eea
The $p$ simplex $\sigma^p$ writes 
\bea
 \Big{\{}
 \cB^{\cC^{n+1}\setminus\{ i_0 \}}_{\cV^{\widehat{i_0}}}, \dots,
  \cB^{\cC^{n+1}\setminus\{ i_p \}}_{\cV^{ \widehat{i_p} }}
 \Big{\}} \qquad \cV\subset \cV^{ \widehat{i_0}} \cap \dots \cap \cV^{ \widehat{i_p}}
\; .
\eea 

The proof is now tautological. For any subset $M^k\subset \sigma^p$ of cardinality $(k+1)<(p+1)$,
\bea
 \Big{\{}
 \cB^{\cC^{n+1}\setminus\{ i_{j_0} \}}_{\cV^{\widehat{i_{j_0}}}}, \dots,
  \cB^{\cC^{n+1}\setminus\{ i_{j_k} \}}_{\cV^{ \widehat{i_{j_k}}}}
 \Big{\}}  \qquad \cV\subset \cV^{ \widehat{i_{j_0}}} \cap \dots \cap \cV^{\widehat{i_{j_k}}}
\eea
with $\{j_0,\dots j_k\}\subset \{0,\dots p\}$ there exists a $k$-bubble obtained by adding all lines
of colors $\{0,\dots p\} \setminus \{j_0,\dots j_k\}$ to $\cV$ and completing the graph thus obtained to a
bubble of colors $ \cC^{n+1} \setminus \Big{[} \{ i_0\dots i_p \} \setminus \{ i_{j_0}\dots i_{j_k} \}   \Big{]}$. 
Consequently $M^k$ is the simplex dual to this bubble, $M^k\in\Delta^{\cG}$, 
and the gluing is a simplicial complex.

\qed

In retrospect one sees that all the link graphs are orientable, as they are allways 
made of colored lines joining vertices of opposite orientation. 
The colored GFT model is the simplest one which guarantees
that all sub simplices, of arbitrary dimension, are allways identified 
coherently in all gluings corresponding to the CGFT 
lines. 

\section{Conclusion}\label{sec:conclusion}

We started out work by a in depth study of singularities in the usual GFT models. We concluded that
highly pathological GFT graphs whose dual gluings are not homeomorphic to pseudo manifolds 
dominate in power counting and proliferate in the perturbative development of the usual GFT's. 
Of course, as long as one analyzes particular examples of ``nice'' graphs one is 
oblivious to this problem. However, the moment one tries to treat the usual GFT's as fully 
fledged quantum field theories and take into account {\it all} the graphs, the pathological 
ones dominate. In order to save the GFT's as quantum field theories and obtain a reasonable
effective behavior one must deal one way or another with this problem.

The solution we present in this paper is to use the Colored GFT mo-dels.
The extra structure encoded in the coloring eliminates the wrapping singularities  
for {\it all} graph and in {\it all} dimensions in a very natural way.

A large amount of work still remains to be done before establishing the colored GFT's 
as quantum field theories, most importantly one would like to find some scaling regime 
in which their effective behavior is dominated by manifold configurations.
On the other hand the language of the colored GFT's could be used as a mathematical 
tool to further the understanding of topology. The encoding of the link graphs into 
the bubbles provides a bridge between topology and combinatorics opening 
up the possibility to obtain, using the latter, new results concerning the former.

\section*{Acknowledgements}

Research at Perimeter Institute is supported by the Government of Canada through Industry 
Canada and by the Province of Ontario through the Mi-nistry of Research and Innovation.

\appendix
\section{Remarks on Simplicial Complexes}\label{app:prfs}

\begin{remark} \label{rem:stlink}
The collection $\text{star}_{\Delta}(v) \setminus \text{lk}_{\Delta}(v) $ is
\bea
 \text{star}_{\Delta}(v) \setminus \text{lk}_{\Delta}(v)  = 
\Big{\{} \{ v\} \cup \tau | \tau\in  \text{lk}_{\Delta}(v)  \Big{\}} \; .
\eea
\end{remark}
\noindent {\bf Proof:} ``$\supset$'': Let any $\tau \in  \text{lk}_{\Delta}(v)$. Then 
$\{v\} \cup \tau \in \Delta$ and $\{v\}\cup \tau \in \text{star}_{\Delta}(v)$. 
But, as $\{ v \} \in \{v\} \cup \tau $ then $\{v\} \cup \tau \notin \text{lk}_{\Delta}(v)$,
thus $\{v\} \cup \tau \in \text{star}_{\Delta}(v) \setminus \text{lk}_{\Delta}(v) $.
 
``$\subset$'': Let $\sigma \in \text{star}_{\Delta}(v) \setminus \text{lk}_{\Delta}(v) $. Then 
   $\sigma \cup \{v\} \in \Delta$ and $v\in \sigma$, thus $\sigma \cup \{v\} =\sigma \in \Delta$. Denote 
$\tau=\sigma\setminus \{v\}$. As $\sigma \in \Delta$, and $\Delta$ is a simplicial complex, $\tau \in \Delta$. 
Therefore $\tau \in \Delta$ , $\tau \cap \{v\} =\emptyset$ and $\tau \cup \{v\}=\sigma  \in \Delta$, hence
 $\tau \in  \text{lk}_{\Delta}(v)$.

\qed

\begin{remark} Let $\Delta$ a simplicial complex with simplices of maximal dimension $n$. For any vertex 
$v$ of $\Delta$
\bea 
\chi(\text{star}_{\Delta}(v) ) =1 \; .
\eea
\end{remark}
{\bf Proof:} The $\text{star}_{\Delta}(v)$ admits the disjoint decomposition
\bea
 \text{star}_{\Delta}(v) = \Big{(}\text{star}_{\Delta}(v) \setminus \text{lk}_{\Delta}(v) \Big{)}
 \cup \text{lk}_{\Delta}(v) .
\eea
Under this decomposition all simplices either belong to 
$\text{star}_{\Delta}(v) \setminus \text{lk}_{\Delta}(v) $ or to  $\text{lk}_{\Delta}(v)$. 

Let any $p\ge 0$ simplex in the star, $\sigma^p$. Either $\sigma^p\in \text{lk}_{\Delta}(v)$ or
$\sigma^p=\{v\} \cup \sigma^{p-1}, \sigma^{p-1} \in \text{lk}_{\Delta}(v)$. Note that by definition the link has
exactly {\it one} $-1$ simplex, namely $\emptyset$, and {\it zero} $n$ simplices.

Denote the number of $p$ simplices in the star $N_p$, and the number of $p$ simplices in the link $n_p$. Then
\bea
 f^{p}\bigl( \text{star}_{\Delta}(v) \bigr)=
f^{p}\bigl( \text{lk}_{\Delta}(v) \bigr) + f^{p-1}\bigl( \text{lk}_{\Delta}(v) \bigr)
\eea
hence, 
\bea
&& \chi(\text{star}_{\Delta}(v))=\sum_{p=0}^{n} (-)^n  f^{p}\bigl( \text{star}_{\Delta}(v) \bigr) \\
&&= \sum_{p=0}^{p} (-)^{p} \Big{(} f^{p}\bigl( \text{lk}_{\Delta}(v) \bigr) + 
f^{p-1}\bigl( \text{lk}_{\Delta}(v) \bigr) \Big{)} =f^{-1}\bigl( \text{lk}_{\Delta}(v) \bigr) =1 \; .
\nonumber
\eea

\qed

\begin{remark}
  The links of a simplicial pseudo manifold $\Delta$ are pure and non branching. 
\end{remark}
\noindent{\bf Proof:} Consider a simplex $\tau^p$ of dimension $p$ in a simplicial pseudo manifold and denote 
$\sigma^n_{(1)}, \dots \sigma^n_{(N)}$ all the n dimensional simplices to which $\tau^p$ is a face.
Denote also $\sigma^{n-p-1}_{(k)}=\sigma^{n}_{(k)}\setminus \tau^p$. 

{\it Step 1:} $\text{lk}_{\Delta} (\tau^p)$  is a (n-p-1) - dimensional pure simplicial complex.

Any $\sigma \in \text{lk}_{\Delta} (\tau^p)$ one has $\sigma\cup\tau \in \Delta$ hence there exists some n dimensional 
simplex $\rho^n \supseteq \sigma\cup\tau$. But $\tau \subseteq \rho^n$ thus $\rho^n=\sigma^n_{(k)}$ for some $k$, and
$\sigma\subseteq \sigma^n_{(k)} \setminus \tau^p = \sigma^{n-p-1}_{(k)}$. 

{\it Step 2:} $\text{lk}_{\Delta} (\tau^p)$ is non branching. 

Let $\sigma^{n-p-2}$ a $(n-p-2)$ simplex in $\text{lk}_{\Delta} (\tau^p)$. It is a face of $r$ simplices
of dimension $(n-p-1)$, $\sigma^{n-p-1}_{(k_1)},\dots \sigma^{n-p-1}_{(k_r)}$. Then $\tau^p\cup \sigma^{n-p-2}$ is
a $n-1$ dimensional simplex of $\Delta$ which is a face of the $r$ simplices 
$\tau^p\cup \sigma^{n-p-1}_{k_i}$ of dimension $n$. As $\Delta$ is non branching, $r=2$ 
hence $\text{lk}_{\Delta} (\tau^p)$ is non branching.

\qed

However the links of simplicial pseudo manifolds are not necessarily stron-gly connected. For example, in the
simplicial pseudo manifold 
\bea
  \Big{\{} \{v,a_1,a_2\},  \{v,a_1,a_3\},  \{v,a_2,a_3\},  \{v,b_1,b_2\},  \{v,b_1,b_3\},  
\{v,b_2,b_3\} \nonumber\\
\{b_1,b_2,m_{12}\}, \{a_1,a_2, m_{12} \} , \{ b_1,a_1, m_{12}\}, \{b_2,a_2, m_{12}\} \nonumber\\
\{b_1,b_3,m_{13}\}, \{a_1,a_3, m_{13} \} , \{ b_1,a_1, m_{13}\}, \{b_3,a_3, m_{13}\} \nonumber\\
\{b_2,b_3,m_{23}\}, \{a_2,a_3, m_{23} \} , \{ b_2,a_2, m_{23}\}, \{b_3,a_3, m_{23}\}  \Big{\}} \; ,
\eea 
the link of $v$ is not even connected, much less strongly connected.

\end{document}